\documentclass[12pt]{article}
\usepackage{graphicx}
\usepackage{amsmath}
\usepackage{bm}
\usepackage{accents}
\usepackage{cite}

\newcommand\pubnumber{}
\newcommand\pubdate{\today}

\def\berkeley{Department of Nuclear Engineering\\
University of California, Berkeley, CA 94720, USA}

\def\Title#1{\begin{center} {\Large #1 } \end{center}}
\def\Author#1{\begin{center}{ \sc #1} \end{center}}
\def\Address#1{\begin{center}{ \it #1} \end{center}}
\def\collaboration#1{\begin{center}{ \it #1} \end{center}}

\newcommand\pubblock{\rightline{\begin{tabular}{l} \pubnumber\\
         \pubdate  \end{tabular}}}
\newenvironment{Abstract}{\begin{quotation}  }{\end{quotation}}
\newenvironment{Presented}{\begin{quotation} \begin{center} 
             PRESENTED AT\end{center}\bigskip 
      \begin{center}\begin{large}}{\end{large}\end{center} \end{quotation}}

\def\beq{\begin{equation}}
\def\eeq#1{\label{#1}\end{equation}}
\def\eeqn{\end{equation}}

\def\beqa{\begin{eqnarray}}
\def\eeqa#1{\label{#1}\end{eqnarray}}
\def\eeqan{\end{eqnarray}}

\let\bar=\overbar

\def\Dslash{\not{\hbox{\kern-4pt $D$}}}
\def\dslash{\not{\hbox{\kern-2pt $\del$}}}

\def\msb{{\bar{\ssstyle M \kern -1pt S}}}

\begin{document}
\begin{titlepage}
\pubblock

\vfill
\Title{HAYSTAC Status, Results, and Plans}
\vfill
\Author{Alex Droster and Karl van Bibber}
\collaboration{For the HAYSTAC Collaboration}
\Address{\berkeley}
\vfill
\begin{Abstract}
We describe the design of the dark matter experiment Haloscope At Yale Sensitive To Axion Cold Dark Matter (HAYSTAC), and report the results of a haloscope search for dark matter axions. We exclude axion models with axion-photon couplings $g_{a\gamma\gamma} \underset{\sim}{>}2\times10^{-14}$ GeV$^{-1}$ over the range $23.55<m_{a}<24.0$ $\mu$eV. This sensitivity is a factor of 2.7 above KSVZ model coupling, averaged over the given mass range. Phase I achieved a noise temperature a factor of 2 over the standard quantum limit. Phase II, now entering commissioning, incorporates a squeezed-vacuum state receiver to evade the quantum limit, which will deliver a factor of $\sim2$ improvement in scanning rate over the current single Josephson parametric amplifier (JPA) receiver. The sensitivity of the HAYSTAC Phase II receiver will lead to precise constraints on two photon coupling strengths and represents the most sensitive axion cavity detector probing $m_a>20~\mu$eV to date.
\end{Abstract}
\vfill
\begin{Presented}
Thirteenth Conference on the Intersection of Nuclear and Particle Physics\\
Palm Springs, CA,  May 29--June 3, 2018
\end{Presented}
\vfill
\end{titlepage}
\def\thefootnote{\fnsymbol{footnote}}
\setcounter{footnote}{0}

\section{\label{sec:level1}Introduction}

Astrophysical measurements over the last few decades overwhelmingly favor a $\Lambda$CDM cosmology in which 85\% of the matter in the universe is in the form of ``dark matter": nonrelativistic, nonbaryonic matter whose exact particle nature is unknown. A prime candidate for dark matter has long been weakly-interacting massive particles (WIMPs)~\cite{wimps}. However, recent attempts to directly detect WIMPs have yielded null results over wide regions of parameter space, disfavoring the WIMP hypothesis and motivating searches for a different type of dark matter particle~\cite{wimps2}.\\
\indent The axion is a hypothetical pseudo-scalar field arising from the Peccei-Quinn solution to the charge parity (CP) violation problem of QCD. An axion in the 1-100 $\mu$eV range is also an attractive candidate for light dark matter~\cite{preskill, abbott, dine}, and recent lattice QCD calculations have favored higher mass axions with mass $m_a>50~\mu$eV~\cite{QCDaxion}. A comprehensive review of the particle physics of the axion, its cosmological and astrophysical significance, and experimental searches for it can be found in Ref.~\cite{AnnRev}.

\subsection{\label{sec:level2}Detection Principle}
\begin{figure}[h]
\centering
\includegraphics[scale=.33]{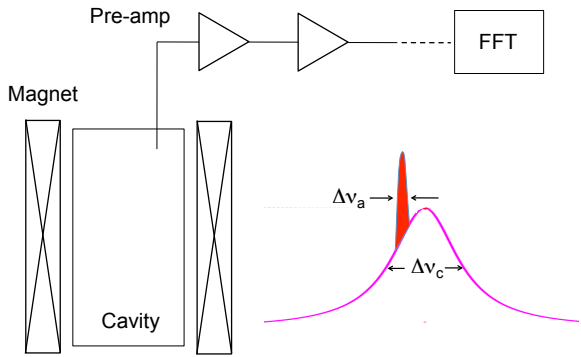}
\caption{\label{fig:axiondetectionschematic}Layout of a microwave cavity axion search experiment. The detector consists of a microwave cavity located inside the bore of a magnet with a highly homogenous magnetic field. Due to virialization, the electric field induced by axion dark matter is not monochromatic but rather has nonzero bandwidth $\Delta v_a$, much less than the cavity bandwidth $\Delta v_c$.
}
\end{figure}
In 1983 Sikivie proposed a novel axion detection strategy, in which a properly tuned microwave cavity in the presence of an external magnetic field could resonantly detect the conversion of a hypothetical axion to a single microwave photon~\cite{sikivie1983}. Because the signal power is proposed to be on the order of $10^{-22}$ Watts~\cite{AnnRev}, operation at cryogenic temperatures and a low noise receiver are necessary for any axion search. Most axion searches to date have used a detection scheme in which a microwave cavity sits in the bore of a superconducting magnet, with power coupled out of the cavity and fed to a low-noise amplifier. A schematic of this axion detection scheme is shown in Fig. \ref{fig:axiondetectionschematic}.\\
\indent The resonant conversion condition is that the axion mass $m_a$ is close to the resonant frequency $\nu_c$ of a cavity mode with an appropriate spatial profile; more precisely, within the linewidth of the mode, $\Delta\nu_c$. Because the axion linewidth is much smaller than the cavity linewidth, the relevant linewidth for a linear receiver is the axion bandwidth rather than the that of the cavity. Improvements to signal to noise ratio may be obtained by averaging the system noise for a time $\tau$, given by the Dicke radiometer equation:
\begin{equation}
SNR = \frac{P_S}{k_B T_S}\sqrt{\frac{\tau}{\Delta \nu_a}}
\end{equation}
where $k_B$ is Boltzmann's constant, $T_S$ is the system noise temperature, and $P_S$ is the signal power exactly on resonance. Assuming a phase-insensitive linear receiver, the system noise temperature is given by
\begin{equation}
k_BT_S=h\nu\Bigg(\frac{1}{e^{h\nu/k_BT}-1}+\frac{1}{2}+N_A\Bigg)
\end{equation}
where the added noise is $N_A\geq\frac{1}{2}$, from which we obtain the standard quantum limit $k_BT_S\geq h\nu$ of cavity axion detection.\\
\indent The signal power is given in natural units by
\begin{equation}
P_S = \Big(g_\gamma^2\frac{\alpha^2\rho_a}{\pi^2\Lambda^4}\Big)\Big(\omega_c B_0^2VC_{mn\ell}Q_L\frac{\beta}{1+\beta}\Big).
\end{equation}
\indent The theory parameters in Eq. (3) beyond the control of the experimentalist are contained in the first term: the coupling constant $g_\gamma$, the fine-structure constant $\alpha$, the local density of axions in the galactic dark matter halo $\rho_a\approx0.45$ GeV/cm$^3$~\cite{jiread}, and the mass of the axion ($m_a$), as encoded in $\Lambda$ ($\Lambda\approx77.6$ MeV).\\
\indent The coupling $g_\gamma$ is a model-dependent dimensionless constant based on the physical coupling that appears in the axion-photon Lagrangian: $g_{a\gamma\gamma}=m_a(g_\gamma\alpha/\pi\Lambda^2)$. Two models, denoted by KSVZ~\cite{kim},~\cite{SVZ} and DFSZ~\cite{DFS}, bound the value of $g_\gamma$ 
and define a ``model band" of cosmologically significant sensitivity.\\
\indent Within experimental control are the frequency of the cavity $\omega_c=2\pi\nu_c$, the applied magnetic field strength $B_0$, the volume of the cavity $V$, the ``form factor" $C_{mn\ell}$ of electromagnetic mode $nm\ell$, and the loaded quality factor of the cavity $Q_L$, the cavity quality factor with power coupled out to the receiver, $Q_L=Q_0/(1+\beta)$.\\
\indent The form factor $C_{mn\ell}$ varies between 0 and 1 and is given by 
\begin{equation}
C_{mn\ell} = \frac{\Big(\int \text{d}^3\mathbf{x}~\mathbf{\hat{z}\cdot E}_{mn\ell}(\mathbf{x})\Big)^2}{V\int \text{d}^3\mathbf{x}~\epsilon(\mathbf{x})|\mathbf{E}_{mn\ell}(\mathbf{x})|^2}
\end{equation}
where $\mathbf{E}_{mn\ell}(\mathbf{x})$ is the normalized electric field of the mode, $\epsilon(\mathbf{x})=1$ is the dielectric constant inside the cavity, where it is assumed that the magnetic field $B$ is in the $\mathbf{\hat{z}}$ direction and uniform. Note that nodes in the electric field lead to cancellations in the numerator, and thus the form factor is only appreciable for low order TM$_{0n0}$ modes.\\



\subsection{\label{sec:level3}HAYSTAC}

\begin{table}[t]
\begin{center}
\begin{tabular}{p{2cm}|p{3.5cm}p{3.5cm}p{2cm}p{2cm}}  
Data Run &  Frequency coverage (GHz) &  Thermal Noise on Resonance, $k_BT_S$ &  
Days axion search & Days rescan\\ \hline
 Run 1  &   5.7-5.8      &    $3h\nu$      &     110 & 23  \\
 Run 2 &  5.6-5.7 &     $2.3h\nu$      &  54 & 53 \\ \hline
\end{tabular}
\caption{\label{table:dataruns}Details for each HAYSTAC Phase I data run.}
\end{center}
\end{table}

\indent Inserting typical values for the Haloscope At Yale Sensitive To Axion Cold Dark Matter (HAYSTAC) Phase I detector into Eq. (2) gives a signal power of $P_S\approx5\times10^{-24}$ W on resonance for a $m_a=24~\mu$eV KSVZ axion with $g_\gamma=-0.97$~\cite{NIM}~\cite{PhysRevLett}.\\
\indent The HAYSTAC Phase I experiment has thus far been used for two data runs: 1) Run 1 covering 5.7-5.8 GHz completed in August 2016 and 2) Run 2 covering 5.6-5.7 GHz completed in July 2017. Run 1 consisted of 110 days of axion search data and 23 days of rescans of possible axion candidates, by achieving noise levels of $k_BT_S\approx3h\nu$ on resonance and $2.2h\nu$ detuning 650 kHz in either direction. Run 2 occurred after cryogenic improvements and consisted of 54 days of axion search data and 53 days of rescanning potential candidates, where 75\% of the rescan time was dedicated to candidates from Run 1. Run 2 achieved average noise levels of $k_BT_S\approx2.3h\nu$ on resonance but with a 40\% reduction in $Q_L$ compared to Run 1. This information is summarized in Table \ref{table:dataruns}.\\
\indent The second generation of the experiment, HAYSTAC Phase II, will exploit two technological advances to achieve the necessary leap in sensitivity: 1) an improved dilution refrigerator and thermal linkage scheme that allows for a lower system noise temperature at no expense to $Q_L$, and 2) a squeezed-vacuum state receiver that improves sensitivity off-resonance thereby allowing for faster scanning speeds. A combination of these advances will deliver more than a factor of two improvement in scanning speed over HAYSTAC Phase I. In Section 2 we discuss the design and results of the HAYSTAC Phase I experiment. In Section \ref{sec:phase2} we discuss HAYSTAC Phase II science goals and the developments necessary to reach these goals.

\section{\label{sec:level4}HAYSTAC Phase I}

\subsection{\label{sec:level5}Experiment Design}
The HAYSTAC detector is sited at Yale University's Wright Laboratory. It consists of a tunable microwave cavity inside the bore of a superconducting solenoid magnet. The magnetic field is highly homogenous and has strength 9 T. The cavity is attached to a gantry anchored to the mixing chamber plate of a dilution refrigerator (DR) which is maintained at $T_C = 127$ mK by a PID (proportional-integral-derivative) controller.\\
\begin{figure*}[h]
\centering
\includegraphics[scale=.4]{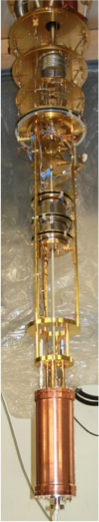}
\includegraphics[scale=.51]{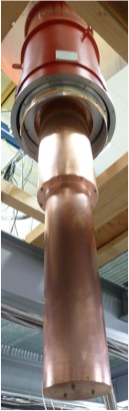}
\includegraphics[scale=.55]{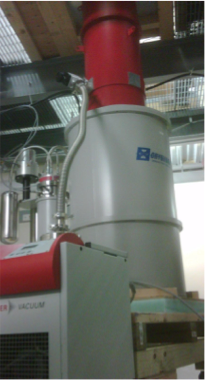}
\caption{\label{fig:gantry}HAYSTAC during integration. \textit{Left}: Gantry, with dilution refrigerator (top), magnetically shielded canister for the amplifier (middle), and microwave cavity (bottom). \textit{Middle}: Thermal shields enclosing the gantry. \textit{Right}: After insertion into the magnet.}
\end{figure*}

\subsubsection{\label{sec:level6}Microwave Cavity}
\indent The microwave cavity is a right circular stainless steel cylinder coated with 0.002 in of oxygen-free high conductivity (OFHC) copper, 10 in length and 4 in diameter (see Fig. \ref{fig:gantry}). Annealing the copper enables the cavity to achieve a near-theoretical value for the quality factor $Q_0$, as limited by the skin depth of the material.\\
\indent The TM-like modes of the cylinder may be tuned by the rotation of a copper plated rod which takes up 25\% of the cavity volume. Rotating this cylinder about an off-center axle provides a continuous tuning range of 3.6-5.8 GHz for the TM$_{010}$-like mode. Typical cavity parameters are $Q_0=3\times10^4$, $C_{010}=0.5$, cavity bandwidth $\Delta \nu_c/\nu_c=Q^{-1}\sim10^{-4}$, axion bandwidth $\Delta \nu_a/\nu_a\sim\beta^2\sim10^{-6}$, and antennae coupling parameter $\beta=2$ to maximize the scan rate~\cite{NIM}.

\subsubsection{\label{sec:level7}Magnet and Cryogenics}

\begin{table}[t]
\begin{center}
\begin{tabular}{l|ccc}  
Data Run &Rod Temperature (mK)&Thermal Noise on Resonance, $T_{\text{sys}}$&$Q_L$\\ \hline
 Run 1    &600&     $3T_{\text{SQL}}$      & $3\times10^4$  \\
 Run 2 &250&     $\sim2T_{\text{SQL}}$     &  $1.8\times10^4$ \\ \hline
\end{tabular}
\caption{\label{table:noisetemps}Total system noise on resonance compared to tuning rod temperature.}
\end{center}
\end{table}

The HAYSTAC dilution refrigerator (DR) was manufactured by VeriCold Technologies\footnote{Oxford Instruments, https://nanoscience.oxinst.com/}. An operating temperature of 127 mK minimizes system noise while at the same time reducing fluctuations observed in JPA gain that occur at lower temperatures~\cite{NIM}. All experimental components share a common vacuum space except for the magnet, which is cooled by its own pulse-tube cryocooler.\\
\indent The magnet\footnote{Cryomagnetics Inc., https://www.cryomagnetics.com} is cooled by a pulse-tube cryorefrigerator and operates in persistent mode. The field is homogenous in the axial ($\mathbf{\hat{z}}$) direction such that the radial component of the magnetic field is less than 50 G in the region of the cavity. In order to minimize the external magnetic field near the superconducting amplifiers, there is an active bucking coil as part of the magnet cryostat, four stategically placed persistent coils, two layers of cryoperm, a layer of lead sheet and a layer of niobium both within the JPA canister. This setup reduces the field at the JPA to less than 1\% of a flux quantum~\cite{NIM}.\\
\indent In order to thermally connect the cavity to the mixing chamber of the DR, a support gantry made of copper alloy clamps the top end cap of the cavity to the bottom of the mixing chamber. The equilibration time for the cavity to respond to a change in temperature in the mixing chamber is on the order of a few minutes. The gantry and DR heat shields are shown in Fig. \ref{fig:gantry}.\\
\begin{figure}[t]
\centering
\includegraphics[scale=.4]{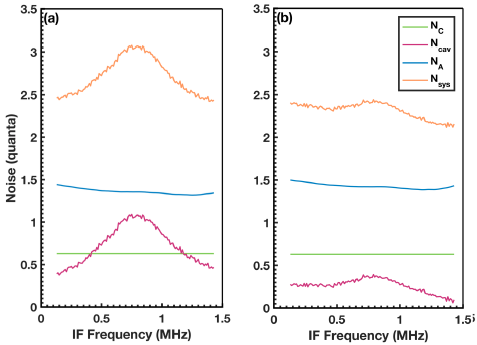}
\caption{\label{fig:adhocmethodgraph}Representative noise measurements from (a) Run 1 and (b) Run 2 after employing an \textit{ad hoc} thermal link. $N_C$ (green) is from thermometry, $N_A$ (blue) is derived from the average off-resonance noise measurements, and $N_{cav}$ (pink) is the excess noise added by the cavity as determined by a single Y-factor measurement during the data run. $N_{sys}$ (orange) is the sum of these contributions.}
\end{figure}
\indent Although the barrel of the cavity was thermally well-sunk to the mixing chamber of the DR, the cavity tuning rod lacked a sufficient thermal connection which resulted in a ``hot rod problem" seen during Run 1~\cite{PhysRevLett}~\cite{PRD}. Prior to Run 2, tests on a similar cavity suggested that insertion of 1/8" diameter copper rods into both ends of the cavity axle could provide sufficient thermal linkage, and this solution was implemented in following data run.  Furthermore, a new piezoelectric tuning mechanism\footnote{Attocube, http://www.attocube.com/} reduced added thermal noise compared to a stepper motor~\cite{PRD}.\\
\indent Fig. \ref{fig:adhocmethodgraph} shows representative noise measurements from each run. Although this solution reduced the average total system noise photon number from 3 quanta in Run 1  to 2.3 quanta in Run 2, the insertion of copper rods into the cavity also caused a 40\% reduction in $Q$ (see Table \ref{table:noisetemps}). This decrease in $Q$ was due to a non-optimal insertion depth of the thin copper rod that resulted in coupling power out of the cavity. HAYSTAC Phase II will solve the hot rod problem with a method that allows the tuning rod to reach base temperature of 100 mK at no expense to $Q$. Section \ref{sec:phase2} details this solution.

\subsubsection{\label{sec:level8}Readout System}

The readout system for HAYSTAC Phase I is based on a Josephson Parametric Amplifier (JPA) to minimize system noise and a microwave switch to enable \textit{in situ} noise calibration. The switch may be toggled between the cavity and a $50~\Omega$ termination anchored on the still plate of the DR at $T_H=775$ mK. Noise calibrations occured once every 11 hours during both axion search data runs.\\
\indent The JPA used for data runs 1 and 2 has a maximum resonant frequency of 6.5 GHz and gain bandwidth product $\sqrt{G}B=26$ MHz. A target gain of $\sim21$ dB is large enough to overwhelm the $\sim20$ quanta of added noise of the high electron mobility transistor (HEMT) amplifier which follows the JPA in the receiver chain but is low enough to avoid deviations from a Gaussian distributed noise spectra that occurs at higher gain.\\
\indent Half an applied DC flux quantum tunes the JPA its full range of 2 GHz. This extreme flux sensitivity necessitates specialized magnetic shielding in the region of the JPA, which is accomplished by a combination of active bucking coils, superconducting coils, and passive components. A full description of the JPA shielding is given in Ref.~\cite{NIM}.\\
\indent Operating the JPA in the phase-sensitive mode offers no improvement in axion search sensitivity because the factor of two improvement in system noise temperature is exactly cancelled by the loss of signal power in one quadrature~\cite{SSRprinciple}. However, a squeezed-vacuum state receiver will be implemented in HAYSTAC Phase II to circumvent the standard quantum limit. This upgrade is described in more detail in Section \ref{sec:phase2}.

\subsection{\label{sec:level9}Phase I Results}
\begin{figure*}[!t]
\centering
\includegraphics[scale=.5]{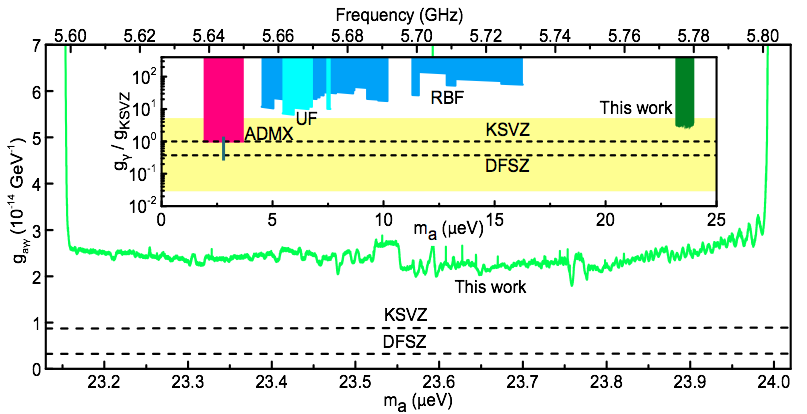}
\caption{\label{fig:exclusionplot}Plotted in green is the HAYSTAC exclusion data at 90\% confidence from data runs 1 and 2. Excluded is the axion mass range $23.15\leq m_a\leq24.00~\mu$eV at coupling $|g_\gamma|\geq2.7\times|g_\gamma^{\text{KSVZ}}|$. This result is the first exclusion of a QCD axion over 20 $\mu$eV. Plotted in the inset are exclusion limits from other axion haloscopes~\cite{hagman, asztalos, asztalos2, asztalos3, sloan, RBF, RBF2, UF}. The cosmologically relevant axion model band is shaded yellow~\cite{chenggeng}.}
\end{figure*}
We report in Fig. \ref{fig:exclusionplot} the 90\% exclusion limit for $g_\gamma$ based on the combined axion search data from runs 1 and 2. Excluded is a two photon coupling of $g_{a\gamma\gamma} \underaccent{\sim}{{>}}2\times10^{-14}$GeV$^{-1}$ over the range $23.55<m_{a}<24.0$ $\mu$eV~\cite{PRD}. This sensitivity is a factor of 2.7 greater than KSVZ model coupling, i.e. $|g_\gamma|\underaccent{\sim}{{>}}2.7\times |g_\gamma^{\text{KSVZ}}|$, and is the first cosmologically relevant exclusion of a QCD axion of $m_a>20~\mu$eV~\cite{PRD}.\\
\indent The 90\% confidence limit was determined by setting a $5.1\sigma$ SNR target, which corresponds to a single bin power excess of $3.455\sigma$ at 95\% confidence for a candidate event. Of the 27 candidate events found in the first scan, none of them exceeded this threshold at 95\% confidence upon rescanning~\cite{PhysRevLett}. Therefore, we claim an exclusion of all bins at 90\% confidence. The number of candidate events, 27, is consistent with that expected based on synthetic Gaussian white noise subject to the same processing~\cite{bensanalysis}.\\
\indent HAYSTAC Phase I is to date the most sensitive axion detector at the $m_a>20~\mu$eV mass range, with total noise a factor of 2 times the standard quantum limit ($T_{\text{sys}}\sim 2T_{\text{SQL}}$). Axion haloscope experiments above a few gigahertz are subject to engineering challenges because the effective volume $VC_{mn\ell}$ drops off rapidly with increasing frequency. Despite this unfavorable scaling, these axion search results demonstrate that a sufficiently low-noise experiment can reach cosmologically relevant sensitivities for this mass range. It is also the first axion experiment to use a dilution refrigerator and JPA.\\
\indent This result concluded Phase I of HAYSTAC. Phase II will include upgrades to cryogenics, an improved double-blind analysis procedure, and the implementation of a squeezed state receiver.

\section{\label{sec:phase2}HAYSTAC Phase II}
The HAYSTAC Phase II instrument design combines new cryogenics and amplifiers to extend the frequency coverage and achieve a factor of 2 improvement in scanning speed over HAYSTAC Phase I. A new dilution refrigerator will increase the cooling power and lower the base temperature of the experiment to 100 mK; an improved implementation of the method for thermalizing the cavity tuning rod will enable lower system noise temperatures at no expense to the cavity quality factor, $Q$; and a new squeezed state receiver will increase the scan speed by a factor of $\sim2$.

\subsection{\label{sec:level11}Cryogenics}
\begin{figure*}[!h]
\centering
\includegraphics[scale=.3]{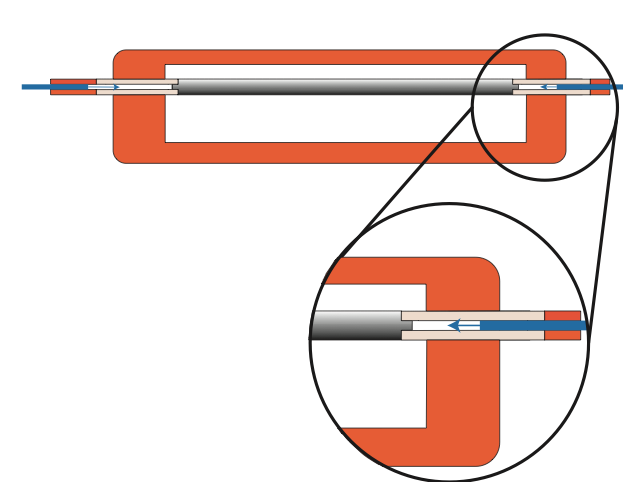}
\includegraphics[scale=.31]{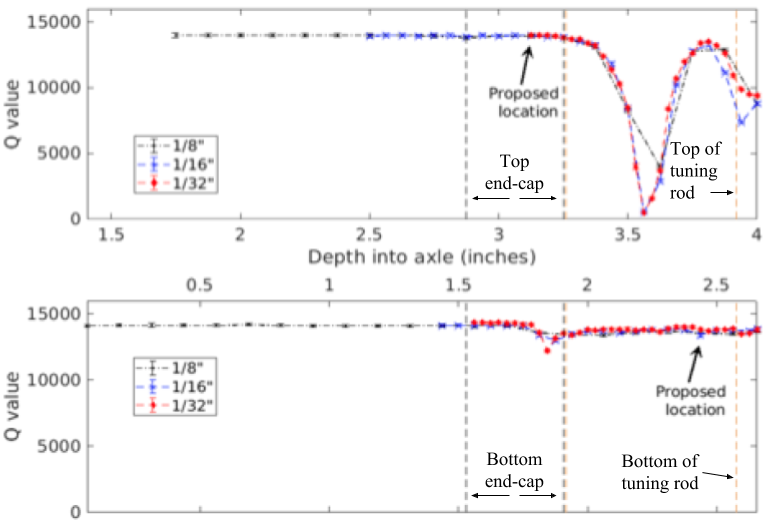}
\caption{\label{fig:thermallink}\textit{Left}: A 1/8" diameter OFHC copper rod (blue) inserted into both ends of the axle completes a strong thermal link between the cavity tuning rod (orange) and the mixing chamber of the DR (not pictured). \textit{Right}: Room temperature tests demonstrate that there exists an insertion depth at which no significant power is coupled out of the cavity. Top and bottom graphs represent data for the top and bottom of the tuning rod, respectively. In the lower graph, this thermalization rod will be inside the cavity, although not making electrical contact with the tuning rod; in the top graph, the second thermalization rod will stop just shy of entering the cavity, but very close to the rod, ensuring adequate thermal contact. Simulations (not shown) suggest that this behavior is not expected to change at cryogenic operating temperature.}
\end{figure*}
The HAYSTAC Phase II cryogenics represents a significant improvement over Phase I. The cavity will be cooled to $\leq100$ mK by a new $^3$He/$^4$He dilution refrigerator\footnote{BluFors Cryogenics, https://www.bluefors.com} that has a factor of 3 improvement in cooling power at base temperature over the Phase I fridge. This new fridge also boasts improved vibration isolation for thermal noise reduction and a new variable temperature stage for \textit{in situ} temperature calibration. A redesigned cavity support structure will mitigate damage in case of a magnet quench.\\
\indent Minimizing system noise temperature requires an improved method of thermalizing the cavity tuning rod to the cold plate of the dilution refrigerator. Two 1/8" diameter OFHC copper rods inserted into the top and bottom of the axle of the tuning rod provides a thermal link between the tuning rod and the rest of the interior of the cryostat. To achieve the necessary thermal connection, we plan to anchor the copper rods to the cavity support structure using copper braids. Fig. \ref{fig:thermallink} shows that at certain insertion depths, the copper rods do not couple significant power out of the cavity, i.e. the insertion of the rods does not affect the cavity quality factor, $Q$. This result is in contrast with the \textit{ad hoc} method of thermalizing the tuning rod which occurred in Phase I, which resulted in a 40\% reduction in $Q$.\\

\subsection{\label{sec:level12}Squeezed-Vacuum State Receiver}
\begin{figure}[!h]
\centering
\includegraphics[scale=.35]{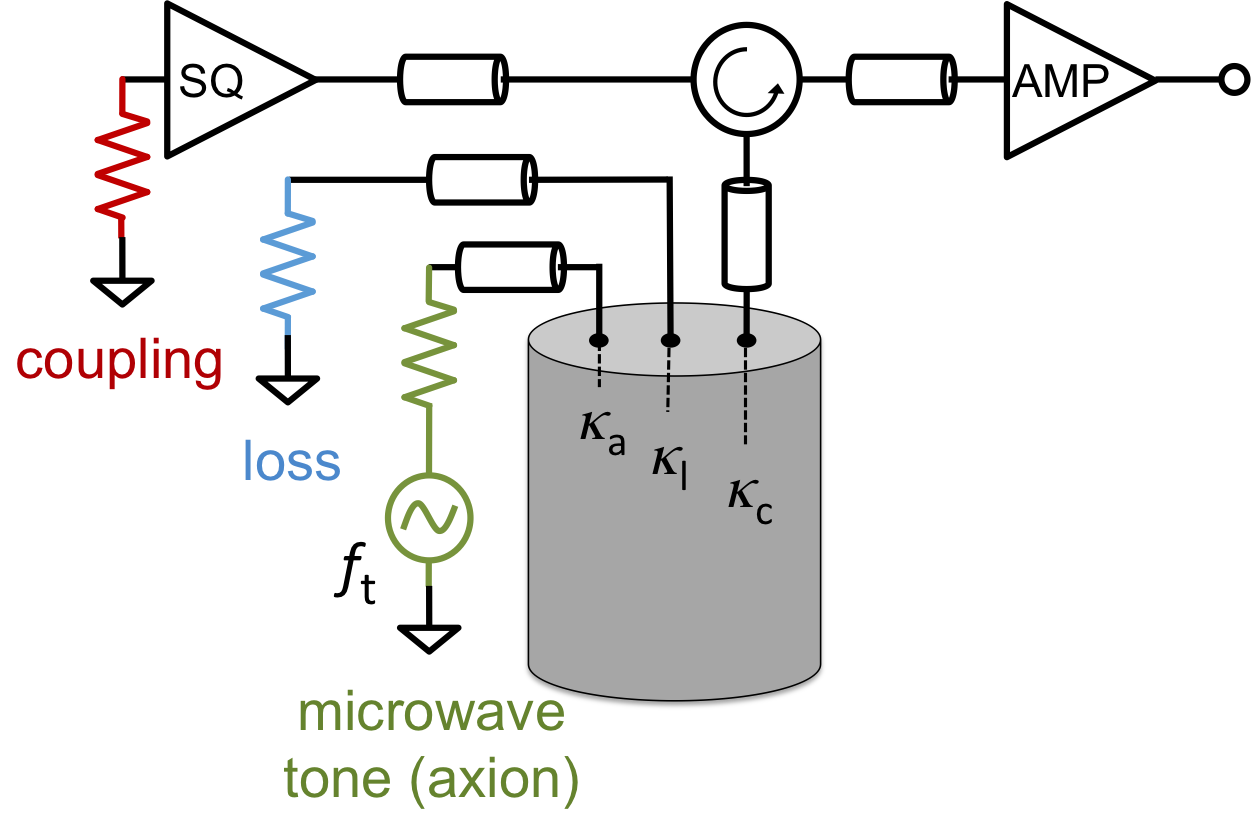}
\caption{\label{fig:ssrschematic}Schematic of a SSR and cavity. In red, a coupling port couples the cavity mode to the mode propagating through a transmission line with power decay rate $\kappa_c$. A microwave circulator separates incoming and outgoing modes. In blue, a fictitious loss port represents represents the cavity's internal power loss with rate $\kappa_\ell$. In green, a fictitious signal port models the coupling of a microwave axion signal to the cavity mode with power loss rate $\kappa_a$.}
\end{figure}

\begin{figure}[t]
\centering
\includegraphics[scale=.5]{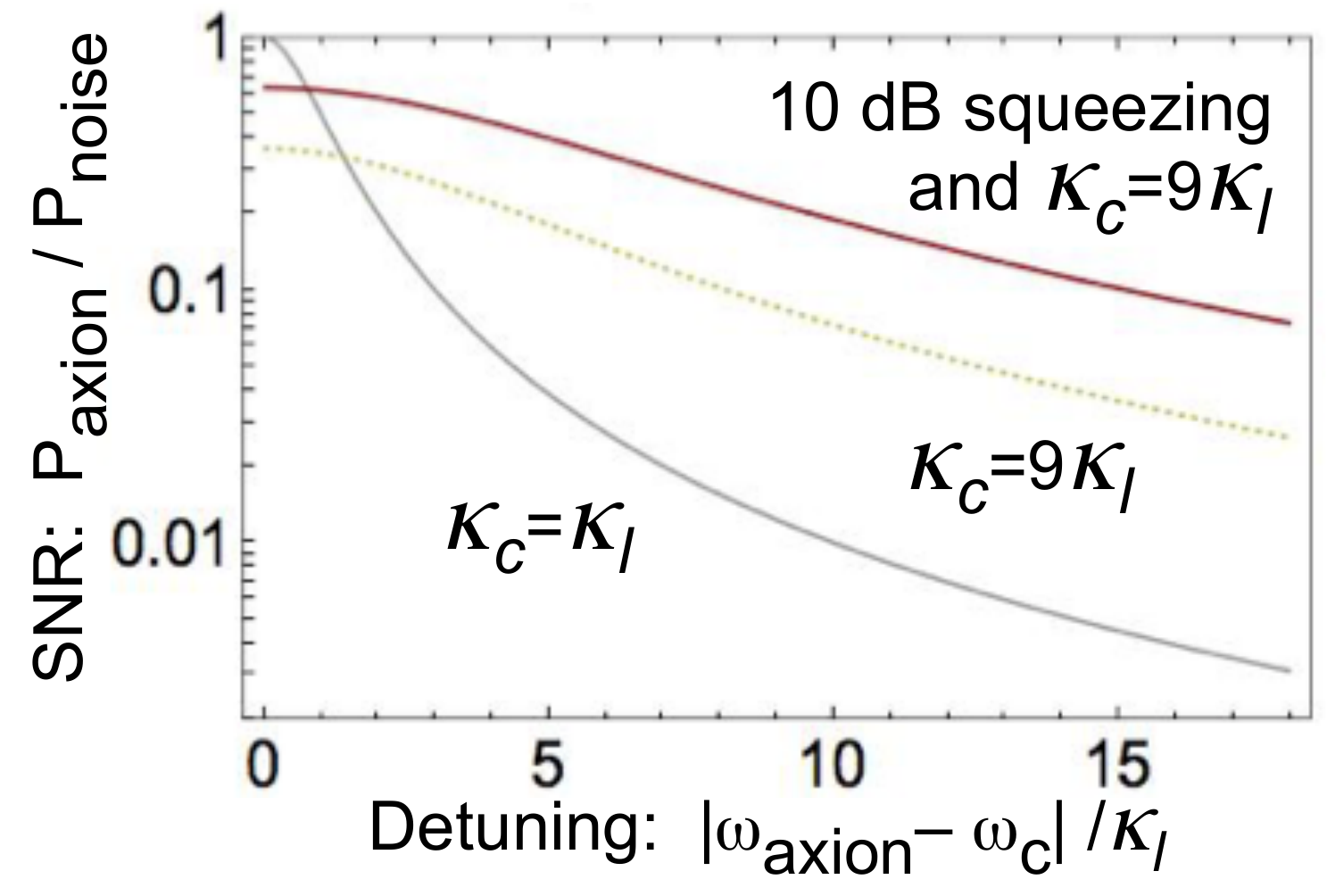}
\includegraphics[scale=.4]{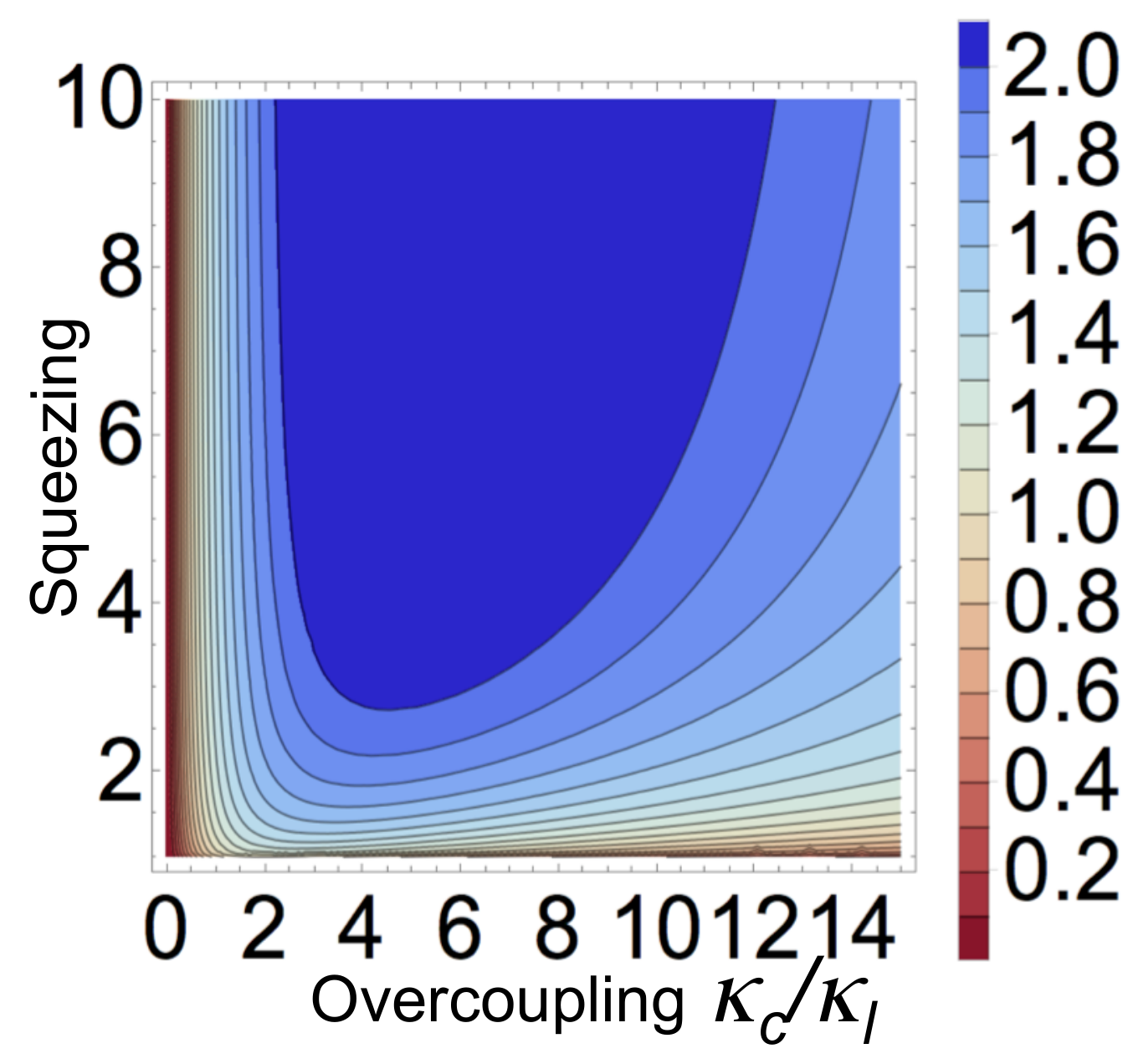}
\caption{\label{fig:scanrategraph} \textit{Left}: Signal to noise ratio as a function of the signal's detuning from cavity resonance ($\omega_c=0$) for three scenarios: 10 dB squeezing at 9$\times$ overcoupling ($\kappa_c=9\kappa_\ell$) (red), 9$\times$ overcoupling with no squeezing (yellow), and no squeezing with critical coupling ($\kappa_c=\kappa_\ell$) (grey). Squeezing does not improve sensitivity for a tone on cavity resonance but rather broadens the bandwidth. \textit{Right}: The scan rate enhancement as a function of gain and coupling ratio. At the measured transmission efficiency of $\eta=0.69$, scan rate enhancement plateaus at 2.2 for 5$\times$ overcoupling.}
\end{figure}

The HAYSTAC Phase II receiver system is based on the implementation of squeezed state receiver (SSR) technology~\cite{caves}, which has been demonstrated at microwave frequencies in numerous proof-of-concept experiments~\cite{murch, clark, bienfait}. Phase I has already demonstrated nominal noise performance, robust detector and JPA setup, and effective cryogenics for reaching model band sensitivity. For HAYSTAC Phase II, improved parameter space coverage requires an increase in scan rate.\\
\indent The SSR consists of two JPAs, denoted SQ and AMP in Fig. \ref{fig:ssrschematic}, which respectively squeeze and amplify a microwave signal interacting with a cavity. The first JPA (SQ) squeezes the vacuum noise entering the cavity through the measurement port (Fig. \ref{fig:ssrschematic}, red), preparing the cavity in a squeezed state along one quadrature of the cavity field, $\hat{X}$, where $\hat{H}=\hbar\omega_c(\hat{X}^2+\hat{Y}^2)/2$ is the Hamiltonian of the cavity mode of interest and $\omega_c$ is the frequency of the same cavity mode~\cite{SSRprinciple}~\cite{ssr}. Inside the cavity, an axion field would appear as a small excess of power in both quadratures. The second JPA (AMP) operated in phase-sensitive mode noiselessly amplifies the $\hat{X}$ quadrature of the measurement port output with enough gain to overwhelm the added noise from the following electronics. A comprehensive review of the use of a SSR in a haloscope axion search can be found in Ref. \cite{ssr}.\\
\begin{figure*}[!t]
\centering
\includegraphics[scale=.484]{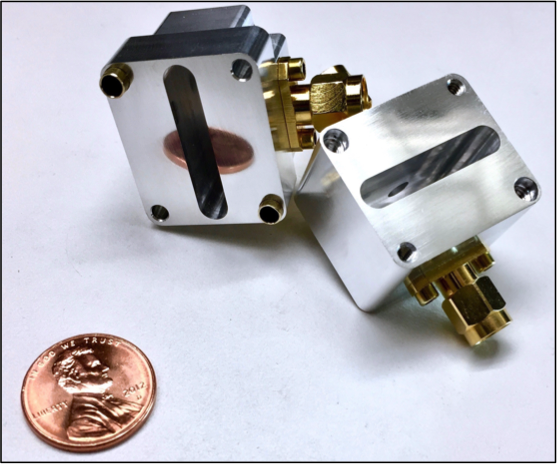}
\includegraphics[scale=.5]{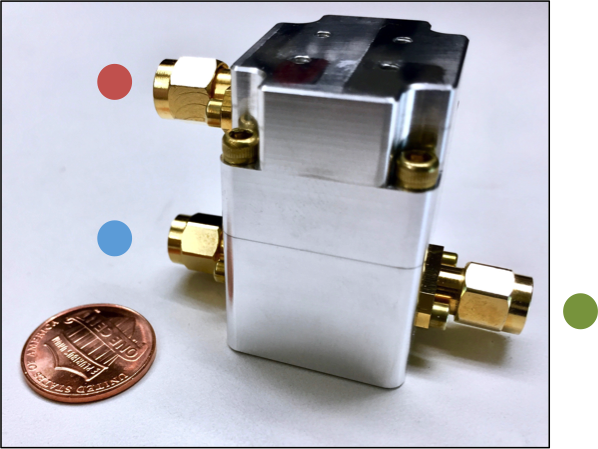}
\caption{\label{fig:ssrphotos}Photos a two-piece 7 GHz testing cavity used in characterization of a squeezed state receiver. Colored dots near each of the microwave transmission ports correspond to the ports in Fig. \ref{fig:ssrschematic} (red for coupling, blue for loss, green for synthetic axion signal).}
\end{figure*}
\indent Fig. \ref{fig:scanrategraph}, left, shows the effect of squeezing on the signal to noise ratio as a function of the signal's detuning from cavity resonance. Overcoupling increases the cavity bandwidth at the cost of reducing sensitivity, but squeezing mitigates this reduction in SNR. Because the axion signal's frequency is \textit{a priori} unknown, this increase in bandwidth at the expense of sensitivity on resonance is a favorable trade-off: the SSR enhances the spectral scan rate for an axion-like signal of unknown frequency by increasing the sensitivity over a broader bandwidth.\\
\indent While increasing squeezing and overcoupling improves the spectral scan rate arbitrarily, in practice transmission losses between SQ and AMP limit the benefit of squeezing because a portion of the squeezed state is replaced with unsqueezed-vacuum~\cite{ssr}. At observed SQ-to-AMP transmission efficiency $\eta=0.69,$ the improvement in scan rate plateaus at
\begin{equation}
\Big(\frac{\text{d}f}{\text{d}t}\Big)_{\text{squeezed}}\Big/\Big(\frac{\text{d}f}{\text{d}t}\Big)_{\text{unsqueezed}}=2.12\pm0.08
\end{equation}
where $f$ is frequency. This enhancement to the scan rate is shown in Fig. \ref{fig:scanrategraph}.  Imperfect transmission efficiency $\eta = 0.69$ limits the maximum scan rate enhancement. Tenfold or greater enhancements require $\eta>0.93$, which may soon be achievable with the development of new quantum technologies such as on-chip circulators and directional amplifiers~\cite{ssr}.

\section{Conclusion}
\begin{figure*}[!h]
\begin{center}
\includegraphics[scale=.4]{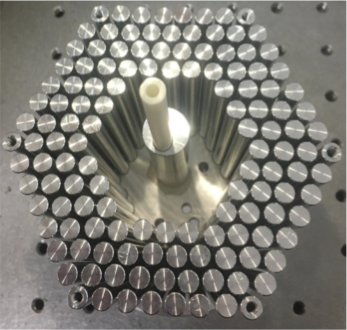}
\includegraphics[scale=.4]{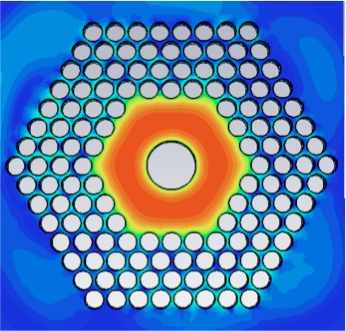}
\caption{\label{fig:pbg} \textit{Left}: Photograph of an aluminum photonic band gap (PBG) structure. A periodic lattice of rods confine TM modes while undesirable TE modes propagate out. \textit{Right}: Simulation\textsuperscript{a} of a PBG structure demonstrates that the defect in the lattice confines the modes of interest (TM modes) while TE modes leak out.}
\end{center}
~~~~\small\textsuperscript{a}CST Microwave Studio, https://www.cst.com/
\end{figure*}
HAYSTAC Phase I reports an exclusion of axion models with $g_{a\gamma\gamma} \underaccent{\sim}{{>}}2\times10^{-14}$ GeV$^{-1}\approx2.7\times|g_\gamma^{\text{KSVZ}}|$ over the range $23.55<m_{a}<24.0$ $\mu$eV and system noise temperature $T_{\text{sys}}\sim 2T_{\text{SQL}}$. The HAYTAC Phase II experiment will be the most sensitive axion dark matter detector operating above $m_a=20~\mu$eV, a development made possible through the advancement of cryogenics and readout technology. This experiment will lead to precise constraints on both the possible mass of axion dark matter particle $m_a$ and possible two-photon coupling constant $g_{a\gamma\gamma}$. To achieve this will require a factor of 2 scanning speed beyond the HAYSTAC Phase I experiment. Phase II will accomplish this goal by a combination of redesigned cryogenics and increased sensitivity off cavity resonance through a squeezed state receiver. Commissioning will begin in late 2018, with the first data run for HAYSTAC Phase II targeted for early 2019.\\
\indent Innovative concepts in both cavity and readout design continue to be explored. A photonic band gap (PBG) resonator promises to eliminate all TE modes interfering with the TM mode of interest (Fig. \ref{fig:pbg}). As the TM mode of interest is tuned in frequency, crossings with untunable TE modes eliminate a significant fraction of frequency coverage. The accelerator physics field has employed PBGs with good results ~\cite{pbg1}~\cite{pbg2}, and they offer a good way to improve frequency coverage should they be adapted to microwave cavity axion searches. Novel readout technology based on single-photon counting methods is also under development. At frequencies greater than 10 GHz, which is not far above the current HAYSTAC search range of 5 to 7 GHz, single-photon detectors become favored when compared to quantum-limited linear amplifiers~\cite{lamoreaux60}. The microwave cavity axion search community may be able to profit from advances in such detection methods based on Rydberg atoms.

\nocite{*}

\bibliographystyle{IEEEtran}
\bibliography{mybib.bib}

\end{document}